\documentclass[traditabstract]{aa}
\usepackage{epsfig}

\usepackage{graphicx,natbib}
\usepackage{natbib} 
\usepackage{amsmath,amssymb}

\def\msun{ \rm M_\odot}
\def\mjup{ \rm M_{\rm J}}
\def\rjup{R_{\rm J}}
\def\mearth{\,{\rm M}_\oplus}

\def\te{T_{\rm eff}}

\def\beq{\begin{equation}}
\def\eeq{\end{equation}}
\def\simgr{\,\hbox{\hbox{$ > $}\kern -0.8em \lower 1.0ex\hbox{$\sim$}}\,}
\def\simle{\,\hbox{\hbox{$ < $}\kern -0.8em \lower 1.0ex\hbox{$\sim$}}\,}

\titlerunning{Structure of the CoRoT exoplanets}
\authorrunning{Leconte et al.}

\begin{document}

\title{Structure and evolution of the first CoRoT exoplanets:\\ Probing the Brown Dwarf/Planet overlapping mass regime}

\author{J. Leconte\inst{1} \and I. Baraffe\inst{1} \and  G. Chabrier\inst{1} \and  T. Barman\inst{2} \and  B. Levrard\inst{1}
}

\institute{ \'{E}cole normale sup\'erieure de Lyon, CRAL (CNRS), 46 all\'ee d'Italie, 69007 Lyon,\\ Universit\'e de Lyon, France (jeremy.leconte,ibaraffe, chabrier@ens-lyon.fr)
\and
Lowell observatory, 1400 West Mars Hill Road, Flagstaff, AZ 86001, USA (barman@lowell.edu)
}

\date{Received 20 February 2009}

\offprints{J. Leconte}


\abstract{
We present detailed structure and evolution calculations for
the first transiting extrasolar planets  discovered by
the space-based CoRoT mission. 
Comparisons between theoretical and observed radii
provide information on the internal composition
of the CoRoT objects.
We distinguish three different
categories of planets emerging from these discoveries and from
previous ground-based surveys: (i) planets
explained by standard planetary models including irradiation, (ii) abnormally bloated planets
and (iii) massive objects belonging to the overlapping mass regime between
planets and brown dwarfs. For the second category, we show that tidal heating can explain the relevant CoRoT objects, providing non-zero eccentricities. We stress that the usual assumption of a quick circularization of the orbit by tides, as usually done in transit light curve analysis, is not justified \textit{a priori}, as suggested recently by Levrard et al. (2009), and that eccentricity analysis should be carefully redone for some observations. Finally, special attention is devoted
to CoRoT-3b and to the identification of its very nature: giant planet or brown dwarf ? The radius determination of this object confirms the theoretical mass-radius predictions for gaseous bodies in the substellar regime but, given the present observational uncertainties, does not allow an unambiguous identification of its very nature.
This opens the avenue, however,
to an observational identification of these two distinct astrophysical populations, brown dwarfs and giant planets, in their
overlapping mass range, as done for the case of the 8 Jupiter-mass object Hat-P-2b. According to the presently published error bars for the radius determination and to our present theoretical description of planet structure and evolution, the high mean density of this object requires a substantial metal enrichment of the interior and is inconsistent at about the 2-sigma limit with the expected radius of a solar-metallicity brown dwarf. Within the aforementioned observational and theoretical determinations, this allows a clear identification of its planetary nature, suggesting that planets may form up to at least 8 Jupiter masses.}


\maketitle

\section{Introduction}
\label{sec:intro}


The first space-based
project devoted to the search for transiting planets, CoRoT, is now collecting
its first results. Seven transiting planets have been announced, with
the most recent one, CoRoT-Exo-7b, 
being a super-Earth of $\sim$ 5-10 $\mearth$, still to be confirmed by radial velocity follow-up. In the present work, presented during the  CoRoT international
symposium taking place in Paris, we analyse the first confirmed transiting planets of CoRoT. They highlight the existence of three different categories of transiting
exoplanets, which have already emerged from the previous ground-based surveys. Such a distinction between three types of
objects is simply used for illustration and does not correspond to different genuine populations of planets.
The first category includes planets whose radii can be explained by standard models and for which a bulk composition can be inferred with little ambiguity. CoRoT-Exo-4b is a template of
this class and  is discussed in \S\ref{sec:sp}. The second kind of objects
concerns inflated planets with abnormally large  radii, which require extra-mechanisms
to explain this behaviour.  CoRoT-Exo-1b and -2b belong to this category and
are analysed in \S\ref{sec:io}. Finally,
the last category of objects, illustrated by CoRoT-Exo-3b,
is populated by massive objects ($\simgr 10\, \mjup$) which lie
in the overlapping mass regime between brown dwarfs and giant planets. 
Special attention is paid to these objects in \S\ref{sec:mso}. We show
how their radius determination can provide valuable information on their true nature and thus on their formation mechanism: gravitational collapse of a molecular cloud for brown dwarfs - yielding objects with the same metallicity as the central star -
against core accretion in a protoplanetary disk for planets (Pollack et al. 1996, Alibert et al. 2005), the most widely accepted scenario for planet formation, yielding a substantially heavy material enriched object, as for our Solar system giant planets. The alternative so-called gravitational instability scenario for planet formation \citep{Bos97} has been shown to be excluded, both on theoretical \citep{Raf05,Raf07} and numerical (\citealt{BDN07}, see \citealt{DDP09} for a recent review) grounds, at least for planets within about 50 to 100 AUs.
In \S\ref{sec:hyp}, we first briefly summarize  the planet structure and evolution models used to
describe the properties of these planets. 


	\begin{table*}[tp]
	\begin{center}
	\resizebox{1\hsize}{!}{
	\begin{tabular}{|c | cccc|c|cc|cccccc|}		
	\hline
& \multicolumn{7}{c}{Planet} & \multicolumn{6}{c}{Star}  \vline \\
&  \multicolumn{7}{c}{}  \vline& \multicolumn{6}{c}{}  \vline \\
Object & M($\mjup$) & R($\rjup$) & Period & a(au) & log($F_{inc}$) & $e$ & $i$ & Spec. Type & $\te (K)$ & Age(Gyr) & M($\msun$) & [Fe/H] & R($R_{\bigodot}$) \\
&  \multicolumn{7}{c}{}  \vline & \multicolumn{6}{c}{}  \vline \\
\hline
&  \multicolumn{7}{c}{}  & \multicolumn{6}{c}{}  \vline \\
CoRoT-Exo-1 b$^a$ & 1.03$\pm0.12$ & 1.49$\pm0.08$ & 1.50 & 0.0254 & 9.467 & - & 85.1 & G0V & 5950 & 0.2-4 & 0.95 & -0.3 & 1.11 \\
CoRoT-Exo-2 b$^b$ & $3.31\pm0.16$ & $1.465\pm0.029$ & 1.74 & 0.0281 & 9.102 & - & 87.84 & K0V & 5625 & 0.2-4 & $0.97\pm0.06$ & & 0.902 \\
CoRoT-Exo-3 b$^c$ & 21.66$\pm1.0$ & 1.01$\pm0.07$ & 4.25 & 0.057  & 9.277 & - & 85.9 & F3V & 6740 & 1.6-2.8 & 1.37 & -0.02 & 1.56 \\
CoRoT-Exo-4 b$^d$ & 0.72$\pm0.08$ & $1.19^{+0.06}_{-0.05}$ & 9.20 & 0.09 & 8.468 & - & 90 & F0V & 6190 & $1.^{+1.0}_{-0.3}$ & 1.1 & & 1.15 \\
HAT-P-2 b$^e$ & 8.04$\pm 0.40$  & 0.98$\pm0.04$  & 5.63 & 0.0677 & 8.962 & 0.5163 & 90 & F8  & 6290 &2.7$\pm1.4$ & 1.31 & 0.12 & 1.48 \\
\hline 
\multicolumn{14}{l}{ $^a$ \cite{BBA08}, $^b$ \citet{AAB08}, $^c$ \citet{DDA08}, $^d$ \citet{MBG08}, $^e$ \citet{WJP07}} \\ 

	\end{tabular}}
	\caption{Transiting Exoplanets characteristics. For all the CoRoT objects, $e=0$ is assumed in the analysis of the light-curve and of the radial velocity data.}
	\end{center}
	\label{table1}
	\end{table*}



\section{Model Description}
\label{sec:hyp}

The main physics inputs used in the present models have been described in previous works devoted to the evolution of brown dwarfs and giant planets \citep{BCB03,BCB08}. 
Evolutionary  models are based on a consistent treatment between non-grey irradiated atmospheric structures and  inner structures. The interior is composed of a gaseous envelope of solar mixture described by the Saumon-Chabrier equation of state (EOS,
\citealt{SCV95}). The inner structure models also take into account enrichment of heavy
material, with the appropriate EOSs \citep{BCB08}. For sake of simplicity, 
the effect of heavy element enrichment is mimicked by the presence of a water core since water, with silicates, is the most abundant species in the protoplanetary disk. \citet{BCB08} show that for enrichment less than 15\% by mass, as 
considered in the present study, the differences in radius between a pure water or a pure silicate core composition, as well as between a heavy material centrally condensed or distributed throughout the envelope, are small ($\lesssim 2\%$). We use the ANEOS EOS for water \citep{TL72}. A detailed analysis
of the effect of the different EOS available in the literature can be found in \citet{BCB08}.
Comparison with models from other groups, in particular by \citet{FMB07}, can also be found in the same paper. 

Transiting planets are very close to their host star ($a< 0.1$AU) and the effect
of stellar irradiation is an important process to be accounted for in their modelling. 
For a 1 Gyr-old Jupiter-like planet orbiting a solar like star at 0.05 AU, the incident stellar flux is more than 10 000 times larger than the thermal flux of the planet itself. In this case, stellar irradiation strongly affects the planet atmospheric structure to deep levels \citep{BHA01}
and thus its evolution (\citealt{BCB03}, \citealt{BSH03}). 
The presence of an irradiated atmosphere acts as a buffer 
which slows down the release of the internal gravo-thermal energy, and
thus the cooling and contraction of the planet \citep{GBH96}. 
We use a grid of irradiated atmosphere models based on the calculations of
\cite{BHA01}, computed for different levels of stellar irradiation relevant
to the present study.


\section{Standard planets}\label{sec:sp}

\begin{figure}[htbp] 
  \centering
  \includegraphics{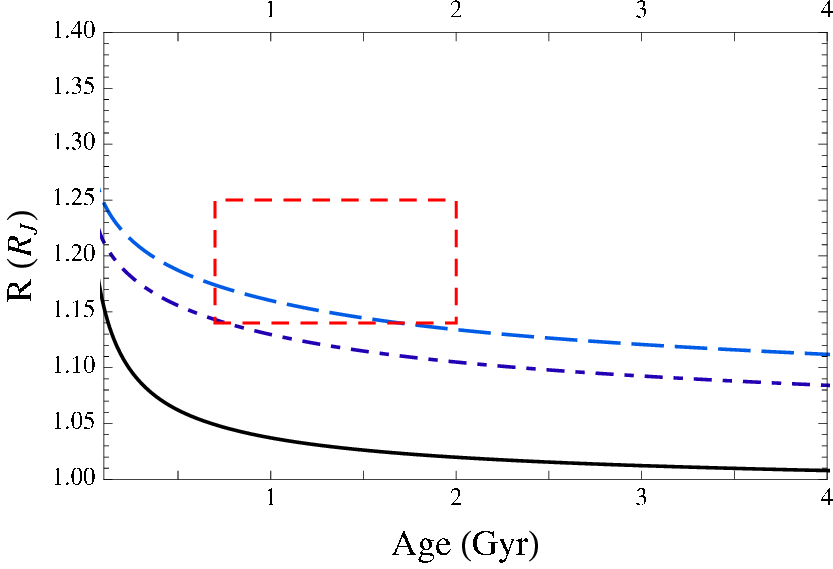}
  \caption{CoRoT-Exo-4b:  evolution of the radius  as a function  of age.
\textit{Solid line}: standard cooling sequence of an isolated 0.72$\mjup$ gaseous sphere with solar composition. \textit{Long-dashed line}: irradiated case. \textit{Dash-dotted line}: Irradiated case with a 10$\mearth$ water core.
\textit{Red box}: observational 1$\sigma$ error bar. }
  \label{fig:c4}
\end{figure}

In this section we focus on planets
whose radius can be explained by the "standard" models discussed in \S\ref{sec:hyp}.
CoRoT-Exo-4b belongs to this population. The planetary and stellar properties of this 
system are given in Table \ref{table1}.  Fig. \ref{fig:c4} shows several evolutionary tracks for CoRoT-4b in a radius versus time diagram.

In order to illustrate the importance of irradiation effects in
the modelling of transiting planets, Fig. \ref{fig:c4} displays
the evolution of an isolated gaseous sphere with solar composition (solid line).
In this case, the predicted radius is too low  compared to the observed
one. 
When taking thermal effects of the stellar impinging flux on the planet's atmosphere into account
(long-dashed line), the model yields a radius consistent with observations. 
For this class of objects, one can infer an upper limit for the content of heavy material
by determining the {\it maximum mass of heavy elements consistent with the lower observational error bar}. In the case of CoRoT-Exo-4b, models with less than 
10$\mearth$ of water remain within the observational error box, as illustrated
by the dash-dotted curve in Fig. \ref{fig:c4}. 
This corresponds to a total mass fraction of heavy elements  $Z\simle 5\%$. As mentioned in \S\ref{sec:hyp},
adopting rock as the main heavy element or a mixture of water and rock will
barely change this value. For this type of transit planet population, the bulk composition can thus be constrained with reasonable accuracy.

One may argue that since a physical process is missing in the modelling of some close-in planets - as discussed in the section below - this upper limit for heavy material enrichment could be underestimated, {\it if} such a process occurs in all transiting planets. An additional heat source, for instance, will yield a larger planet's inflation at a given age, allowing for a larger maximum amount of heavy material consistent with the observational error box.
The exact nature of such additional heating mechanisms (discussed in \S\ref{sec:io}), or of other physical mechanisms invoked to explain the abnormally large radii, however, is still debated and
may depend on the
planet/star system parameters. Under such conditions, the wisest (and simplest)
assumption is not to invoke missing physics when it is not required. One should bear in mind, however, that, given the remaining limitations in our present understanding of planet structure and evolution, these heavy element enrichment determinations retain some degree of uncertainty. This \textit{maximum mass} should rather be seen as the maximum enrichment consistent with the observed $1\sigma$ error bar, according to the present theoretical models.



\section{Abnormally inflated planets}\label{sec:io}

\begin{figure}[htbp]
  \centering
  \includegraphics{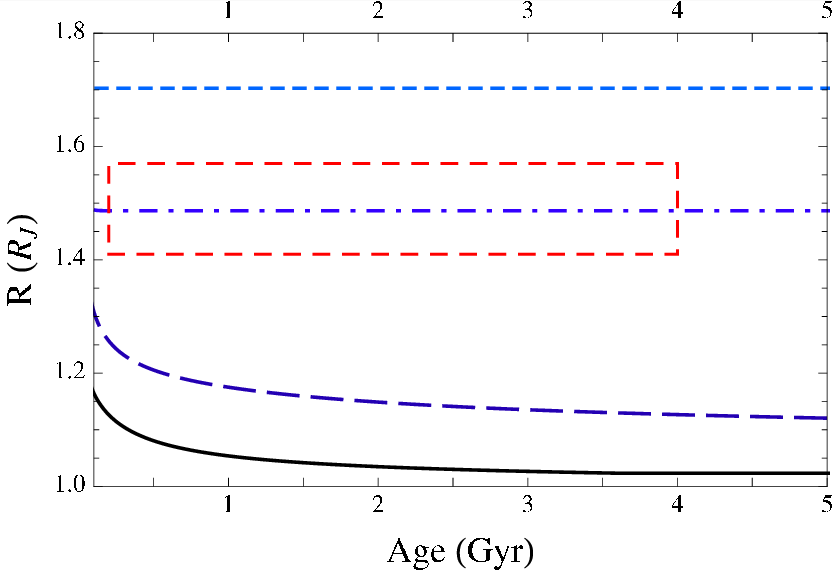}
  \caption{CoRoT-Exo-1b. \textit{Solid line}: isolated solar composition giant planet.\textit{ Long dashed line}: irradiated case. \textit{Dash-dotted line}: effect of tidal heating computed with $e=0.02$. \textit{Short dashed line}: solar composition case with 1\% of the stellar flux transported deep into the planet interior .}
  \label{fig:c1}
\end{figure}

\begin{figure}[htbp]
  \centering
  \includegraphics{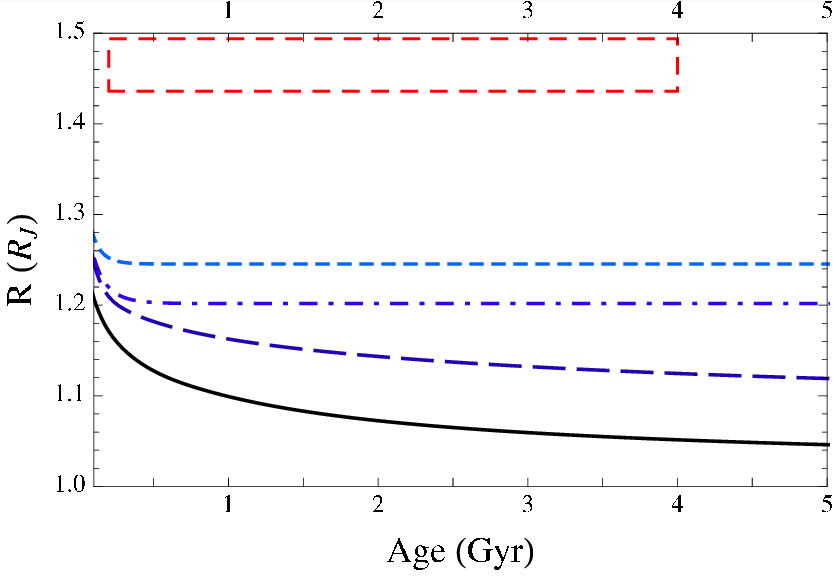}
     \caption{CoRoT-Exo-2b. Same legend as in Fig. 2.}
  \label{fig:c2}
\end{figure}

A significant fraction of transiting planets shows abnormally large radii \citep{MOC07}
compared to  predictions of standard irradiated models (see e.g. \citealt{Gui08};  \citealt{CBL08, BCB09} and references therein). So far, two objects in the CoRoT sample belong to this category :
CoRoT-Exo-1b and -2b. Comparison between models and observations for these
two planets
are shown in Fig. \ref{fig:c1} and \ref{fig:c2}, respectively. 
The effect of irradiation alone does not allow to reproduce the observed radii of these strongly inflated objects. 
These observations thus confirm  the existence of a missing  mechanism in the models, which slows down the cooling
and the contraction of the planet.

Several possibilities 
have been suggested  to explain this puzzling property. Tidal heating due to circularization of the orbit, as originally suggested \citep{BLM01} and then rejected on the basis of a too short characteristic timescale compared with the age of the systems, might provide or at least participate to
the lacking mechanism in some cases, provided tidal effects in the planet and in the {\it star} are properly taken into account (Jackson et al. 2008) \citep{LWC09,IB09}. In a recent paper, however, \cite{LWC09}
show that most transiting systems are not in a state of tidal equilibrium and thus that tidal circularization timescale estimates based on equilibrium tides are not correct. A major consequence of \cite{LWC09} calculations is that assuming zero-eccentricity, when not directly determined, in transiting
light curve analysis, is not necessarily correct and such analysis should be redone carefully.
Tidal heating might thus in some cases provide an extra source of energy during a significant fraction of the planet's evolution.
In the present models, we account for this source of energy,
according to \cite{Hut81}. Note that, since zero eccentricity is assumed 
for the analysis of all the CoRoT transit light-curves, we 
assume a small but finite value for $e$ in order to have a quantitative estimate of
the effect of tidal heating on the bloated CoRoT planets.
In the case of CoRoT-Exo-1b, an eccentricity $e=0.02$ provides enough tidal heating
to reproduce the observed radius within the error bars (dash-dotted curve in Fig. \ref{fig:c1}). For CoRoT-Exo-2b, an eccentricity of only a few percents does not provide
enough tidal heating and larger values are required to reproduce the
radius. We find that a value $e=0.15$ yields a radius in agreement
with observations. 

\cite{SG02} suggested another heating mechanism in the deep interior
of planets, originating from the strong winds generated at the planet's surface. Their numerical simulations of atmospheric circulation produce a downward kinetic energy flux 
of about 1\% of the absorbed stellar incident flux, which dissipates in the interior 
and slows down the planet's contraction \citep{GS02}. Although the validity of this scenario is still debated, with various simulations producing different results (see e.g. \citealt{SMC08}), it is worth exploring this issue.
To test this effect, we have included an extra source of energy corresponding to 1\% of the stellar 
impinging flux contribution in our evolutionary models, as done in \cite{CBB04}. In the case of 
CoRoT-Exo-1b (dashed curve in Fig. \ref{fig:c1}), the resulting theoretical radius
is too large, by $\sim$ 8\%, compared to the observed value, while for CoRoT-Exo-2b, the same relative contribution yields too small of a radius. Note that 
for CoRoT-Exo-4b, analysed in \S  \ref{sec:sp}, models including
1\% of the stellar  impinging flux as an extra source of energy also overestimate by 4\% the radius compared
to the observed value, except if the heavy element content of the planet is significantly larger than the inferred $\sim 5\%$ mass fraction. 
Enhanced atmospheric opacity in transiting planets \citep{BHB07}, although not excluded, remains so far too much of an ad-hoc suggestion to be examined in detail (see \citealt{BCB09} for a discussion). At any rate, the presently detected most inflated transiting planets, like Tres-4b and WASP-12, can not be explained even with such enhanced-opacity models \citep{Gui08,BCB09}.

An other mechanism, based on (inefficient) layered or
oscillatory convection in some planet interiors, has been suggested by \cite{CB07} and has been shown to possibly explain the abnormally large radii.
If future follow-up observations of CoRoT-Exo-1b and CoRoT-Exo-2b confirm
eccentricity values $e< 0.01$, such a mechanism will have to be considered with serious attention. 



\section{Massive substellar objects}\label{sec:mso}

The discovery of "super" Jupiters, 
with masses $\simgr 10 \mjup$, in close orbit to a central star,
raises questions about their nature: planet or brown dwarf ? 
CoRoT-Exo-3b (see Table \ref{table1}) is a perfect example
of such an  ambiguity.  Studies of low mass 
stars and brown dwarfs in young clusters suggest a continuous mass function down to 
$\sim 6\,\mjup$ \citep{CBR07}, indicating that the same formation process
responsible for star formation can produce objects down to a few Jupiter masses.
Indeed, analytical theories of star formation (\citealt{PN04}, \cite{HC08}) show that gravoturbulent fragmentation of molecular clouds
produces, with the same underlying processes, stars and brown dwarfs down to a few Jupiter-masses in
numbers comparable to the observationally determined distribution.   
Brown dwarfs and planets thus overlap in mass, stressing the need for
identi?cation criteria enabling the distinction between these two
types of astrophysical bodies.
The presence of strongly non-solar atmospheric abundances,
as observed in the atmosphere of the giant planets of our Solar System, may provide  signatures of a planetary formation process in a proto-planetary disk. 
Such a signature, however,  is difficult both to observe and to characterize at the present time \citep{CBS07} and may not apply to irradiated planets, with radiatively stable outer layers. A more robust signature of the planet formation process, as expected
from the core accretion model, is the presence of a significant amount of
heavy material in the interior. Observed radii signi?cantly {\it smaller} than predicted for solar or nearly-solar metallicity objects reveal the presence of such a signi?cant average amount of heavy material; a major argument in favor of the core-accretion planet formation process. 
On the opposite, if a physical mechanism is missing in current planet cooling models, as discussed in \S\ref{sec:io}, observed radii {\it larger} than predicted do not necessarily imply an absence or a small amount of heavy material. For such cases, the nature of the object remains ambiguous, if only based on the knowledge if its mean density. 


\begin{figure}[htbp]
  \centering
  \includegraphics{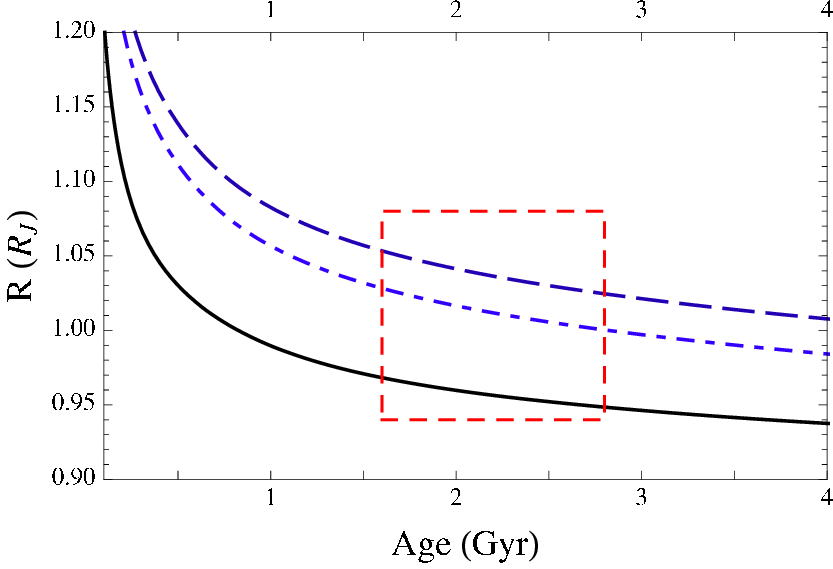}
  \caption{CoRoT-Exo-3b. \textit{Solid line}: standard cooling sequence of an isolated 21.66$\mjup$ brown dwarf with solar composition. \textit{Long dashed line}: irradiated case. \textit{Dash-dotted line}: Irradiated case with a 272$\mearth$ core of water.
\textit{Red box}: observational 1$\sigma$ error bar.}
  \label{fig:c3}
\end{figure}

In this section, we focus on the case of CoRoT-Exo-3b and examine whether 
its radius determination enables us to identify its very nature. 
As shown in Fig. \ref{fig:c3}, the observed radius of CoRoT-Exo-3b can be matched by the model of an {\it irradiated} brown dwarf of $21.6\,\mjup$ with solar composition (long-dashed line). This is by itself an  encouraging confirmation of the theoretical prediction of the age-mass-radius relationship 
in the brown dwarf regime \citep{CB97,BCAH98}.
Note that, given the small orbiting distance, the  effects of irradiation are not negligible, even for such massive objects.
Accounting for irradiation on the atmospheric profile, and thus on the object's cooling history, is thus mandatory to provide consistent comparison between models
and observations, when the radius is determined at this level of accuracy ($\sim 7 \%$).
The present radius error bars, however, are still too large to infer or exclude
the presence of a significant amount of heavy material in the interior of this object.
As done in \S\ref{sec:sp}, a maximum amount of heavy material
can be determined, for the minimum theoretical radius allowed by the error box.
We find an upper mass limit for the core of about 800$\mearth$, i.e. a global maximum mass fraction $Z\lesssim 12\%$. We stress that this corresponds to the {\it maximum} enrichment compatible with the actual error bars. The possibility to 
have such a large amount of heavy material must be examined in the context of our current understanding of planet formation, within the framework of the core-accretion model. Following our previous analysis (\cite{BCB08} and references therein), we can estimate the maximum amount of
heavy material available in the proto-planetary disk for planet formation.
According to current models of planet formation which include migration \citep{AMB05}, up to $\eta \sim 30\%$ of heavy elements contained in the protoplanetary disk can be incorporated into forming giant planets (\citealt{MAB08,MAB09}). 
The maximum mass of available heavy material that can be accreted to form planets is thus:
\begin{equation}
M_Z\approx \eta  \cdot Z \cdot (f \cdot M_\star)
\end{equation}
where $f\cdot$ $M_\star$ is the maximum mass for a stable disk ($\lesssim0.1 M_\star$) and $Z$ is the metal mass fraction of the star.
For CoRoT-Exo-3b, which is orbiting a 1.37$\,\msun$ F star with near solar metallicity, at most $M_Z\approx 270 \mearth$ of heavy material can thus be accreted to form the planet. This (admitedly crude) upper limit derived from current planet core accretion formation models yields a planet contraction consistent with today's observations, as seen in Fig. \ref{fig:c3}. Note that, as discussed in \citet{BCB08} for HAT-P-2b, this heavy material does not need to be accreted into one single object, as very massive planets, in particular short-period ones, may result from smaller planet collisions. Therefore, given the present
uncertainties in the radius determination, neither the brown dwarf nor the planet possibility
can be assessed or excluded for CoRoT-Exo-3b, whose nature remains ambiguous.
A comparison between the predicted radius of a (irradiated) solar-metallicity brown dwarf (dashed line) and of a planet with the aforedetermined massive core, which represents only a $\sim 4\%$ metal enrichment (dash-dotted line) in Fig. \ref{fig:c3}, shows that
a radius accuracy  $\simle 3\%$ is required to resolve the ambiguity, according to the present models. In any event, this demonstrates the promising powerful diagnostic provided by mass-radius determinations to distinguish massive planets from low-mass brown dwarfs, providing adequate observational accuracy.

\begin{figure}[htbp]
  \centering
  \includegraphics{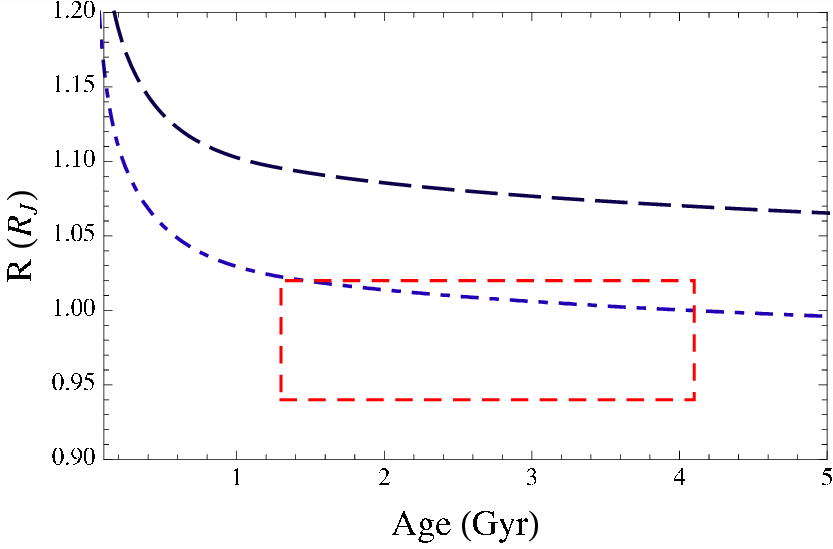}
  \caption{Hat-P-2b. \textit{Long dashed line}: Cooling sequence of an irradiated 8.04$\mjup$ brown dwarf. \textit{Dash-dotted line}: Irradiated case with a 340$\mearth$ core.
\textit{Red box}: observational 1$\sigma$ error bar.}
  \label{fig:hat}
\end{figure}

Among the few known massive planetary-mass objects, there is
one example for which such a radius measurement provides the identification of its nature.
This is the case of Hat-P-2b, a 8 $\mjup$ mass object, closely orbiting an
F type star (see \citealt{WJP07} and Table \ref{table1}).
As illustrated in Fig. \ref{fig:hat},  an irradiated brown dwarf model
(long-dashed line) overestimates the radius by $\sim$ 5\%.
Models including a 340$\mearth$ core mass\footnote{Note that - as discussed in \S \ref{sec:hyp} - this amount of heavy material does not necessarily need to be in a core but could be distributed all over the planet.} can explain the measured radius (dash-dotted line) . Although this amount of heavy material is about the limit of what is available for planet formation, according to current core-accretion models (as estimated from Eq. (1) for the HAT-P-2 system;
see also \citet{MAB09}), the presence of such a metal enrichment ($Z\lesssim15\%$) provides the simplest plausible explanation for the observed radius of HAT-P-2b, according to the present theory. As mentioned earlier, this 340$\mearth$ core for HAT-P-2b should be seen as a rough estimate of the upper limit for the available heavy material in the system, but this analysis shows that the currently observed \textit{low} radius of this object cannot be explained without a substantial enrichment.
It would certainly be interesting to see whether planet models from other groups yield or not similar determinations. Given the fact that these various planet models share many common physics inputs (in particular the H/He and heavy element EOS), it would be surprising that they reach severely different conclusions. While keeping in mind the remaining uncertainties in planet cooling theory, the present analysis provides - with the parameters observed so far - a confirmation of the validity of the core-accretion model, and makes Hat-P-2b the first confirmed \textit{8 $\mjup$ genuine planet} formed by core-accretion in a proto-planetary disk.


\section{Conclusion}\label{sec:disc}

This work focusses on the modelling of the first confirmed transiting planets discovered
by the CoRoT mission. We have distinguished three sorts
of objects. First,
planets whose radius can be explained by standard structure and evolution planetary models including 
the effect of irradiation from the parent star. For these objects, such as CoRoT-Exo-4b,
there is no need to invoke extra physical mechanisms and, in that case, an upper limit can be inferred on their global heavy material content. The second category of objects is characterised  by an abnormally large radius, with two examples in the current CoRoT sample.
Ground-based surveys already found a significant fraction
of such planets and CoRoT confirms this trend. We show that a small but finite
eccentricity of $\sim$ 0.02 
provides enough tidal energy dissipation to explain the radius
of CoRoT-Exo-1b. For CoRoT-Exo-2b, a significantly larger value, $e \sim 0.15$, is
required. We emphasize the fact that a zero eccentricity value is {\it assumed} in the light curve and radial velocity data analysis of current CoRoT planets, and of many other transiting objects. This
hypothesis is based on the idea that (i) tidal circularisation of the orbit is the asymptotic equilibrium state of these planets, (ii) this circularisation
occurs on short timescales
compared to the system's age. Two assumptions which have recently been shown not to be necessarily
correct (\cite{LWC09}). Eccentricity determinations of transiting planets, when unknown, should thus be redone in light of these results.
Our estimates for the required eccentricity to lead to enough tidal dissipation to explain the observed radii
could thus be compared with future follow-up observations/determinations.

The third category of planets includes the "massive" object 
CoRoT-Exo-3b, in the overlapping mass regime between brown
dwarfs and giant planets. We show that the remaining large uncertainties in the radius
determination ($\sim 7 \%$) for this object do not allow a clear identification of its nature. For this object's mass,
present models predict less than 3\% difference between the radius of a brown dwarf (solar composition) and of a planet with a realistic  heavy material enrichment. By itself, the agreement between the observed radius and the theoretical predictions is a beautiful confirmation of the validity of these latter in the brown dwarf mass range, and brings confidence in the validity of the physics included in these models. Although our analysis is inconclusive for CoRoT-Exo-3b, given the present radius uncertainties, we show that the radius determination is discriminant in the case of Hat-P-2b, which is thus the first confirmation of the possibility to form
massive planets  by core accretion (possibly with subsequent collisions) up to $m \simgr 8\, \mjup$. Finally, our analysis shows that, according
to the present models, a typical $\lesssim$ 5 \% accuracy on the radius determination must be achieved in future space-based or ground based transit detections to clearly distinguish planets from brown dwarfs in their overlaping mass domain. We stress, of course, that the conclusions of the present paper are based on presently published observational error bars and on our present theoretical description of planet structure and evolution, still hampered by many uncertainties. The present results thus need to be confronted to e.g. future COROT or KEPLER detections in order to assess their reliability. 

\bibliography{biblio} 
\bibliographystyle{aa}


\end{document}